\documentclass[a4paper]{jpconf}
\usepackage{graphicx}
\begin{document}
\title{Magnetization plateaux in an extended Shastry-Sutherland model}

\author{Kai Phillip Schmidt}

\address{ Lehrstuhl f\"ur theoretische Physik, Otto-Hahn-Stra\ss e 4,\\D-44221 Dortmund, Germany}

\ead{schmidt@fkt.physik.tu-dortmund.de}

\author{Julien Dorier}

\address{Institute of Theoretical Physics, \'{E}cole Polytechnique F\'{e}d\'{e}rale de Lausanne, CH-1015 Lausanne, Switzerland}

\ead{julien.dorier@epfl.ch}

\author{Frederic Mila}

\address{Institute of Theoretical Physics, \'{E}cole Polytechnique F\'{e}d\'{e}rale de Lausanne, CH-1015 Lausanne, Switzerland}

\ead{frederic.mila@epfl.ch}

\begin{abstract}
We study an extended two-dimensional Shastry-Sutherland model in a magnetic field where besides the usual Heisenberg exchanges of the Shastry-Sutherland model two additional SU(2) invariant couplings are included. Perturbative continous unitary transformations are used to determine the leading order effects of the additional couplings on the pure hopping and on the long-range interactions between the triplons which are the most relevant terms for small magnetization. We then compare the energy of various magnetization plateaux in the classical limit and we discuss the implications for the two-dimensional quantum magnet SrCu$_2$(BO$_3$)$_2$.     
\end{abstract}

\section{Introduction}

The two-dimensional frustrated quantum magnet SrCu$_2$(BO$_3$)$_2$ displays a fascinating sequence of magnetization plateaux\cite{onizuka00,kodama02}, which has triggered a lot of activities. Most theoretical studies have been devoted to the properties of a 2D spin-1/2 Heisenberg model 
known as the Shastry-Sutherland model\cite{shastry82} in a magnetic field. In the parameter regime relevant for SrCu$_2$(BO$_3$)$_2$, the ground state of this model is exactly given by the product of dimer singlets\cite{miyahara99}, and the elementary excitations of the model are elementary triplets (triplons\cite{schmi03}). The magnetization process can then be described in terms of polarized triplons having a magnetic quantum number $S_z = +1$. The polarized triplons are hardcore bosons which interact and move on an effective square lattice\cite{momoi00,miyahara03R}. The competition between the kinetics and the interaction of the triplons can lead to a rich phase diagram since a finite magnetic field creates a finite density of triplons\cite{rice02}. For the case of the Shastry-Sutherland model it is known that the interaction between the triplons is the dominant part which is a consequence of the strong frustration. A strong interaction leads to the formation of magnetization plateaux which correspond to Mott insulating phases, i.e. triplons are frozen in the ground state of the system in a regular pattern.

All theoretical approaches agree on the presence of magnetization
plateaux at 1/3 and 1/2\cite{momoi00,miyahara03R,miyahara00,misguich01,miyahara03}, in agreement with experiments\cite{onizuka00,sebastian07}. However, the structure below 1/3 is rather controversial. On the experimental
side, the original pulsed field data have only detected two anomalies interpreted as
plateaux at 1/8 and 1/4\cite{onizuka00}, but the presence of additional phase transitions and of
a broken translational symmetry above the 1/8 plateau has been established by recent
torque and NMR measurements up to 31 T\cite{takigawa07,levy08}. The possibility of additional plateaux has been pointed out by Sebastian et al\cite{sebastian07}, who have interpreted their high-field torque measurements
as evidence for plateaux at $1/q$ with $2\le q\le 9$ and at $2/9$.
On the theoretical side, the situation is not settled either. The finite clusters 
available to exact diagonalizations prevent reliable predictions for high-commensurability
plateaux, and the accuracy of the Chern-Simons mean-field approach
initiated by Misguich {\it et al.}\cite{misguich01} and recently used by Sebastian {\it et al.}\cite{sebastian07} to explain additional plateaux is hard to assess. The essential difficulty lies in the fact that, since plateaux 
come from repulsive interactions between triplons, an accurate determination of the low-density,
high-commensurability plateaux requires a precise knowledge of the long-range part of the
interaction. 

The long-range part of the interaction has been determined recently by perturbative continuous unitary transformations(PCUTs)\cite{dorier08}. It has been found that only a few two-body interactions are decisive at small values of the magnetization and a surprising sequence of plateaux at magnetizations $1/9$, $2/15$, and $1/6$ has been deduced by analyzing the effective model in the classical limit\cite{dorier08}. This has the advantage that almost all commensurabilities can be considered. Note that the effective model obtained by PCUTs is in remarkable accordance with a recent contractor renormalization (CORE) calculation\cite{abendschein08}. 

The obvious contradictions between current theoretical calculations for the magnetization curve in the Shastry-Sutherland model and experimental investigations of the compound SrCu$_2$(BO$_3$)$_2$ certainly opens the question whether the theoretically considered Shastry-Sutherland model is sufficient or if additional magnetic couplings are necessary. A recent ab initio calculation\cite{mazurenko08} has suggested possible extensions of the Shastry-Sutherland model which can be either SU(2) invariant couplings of longer range (in-plane or intra-plane) or Dzyaloshinskii-Moriya (DM) interactions. In this work we will study the leading order effects of the two dominant SU(2) invariant in-plane couplings.    

\section{Model}

\begin{figure}[h]
\begin{center}
\includegraphics[width=0.9\textwidth]{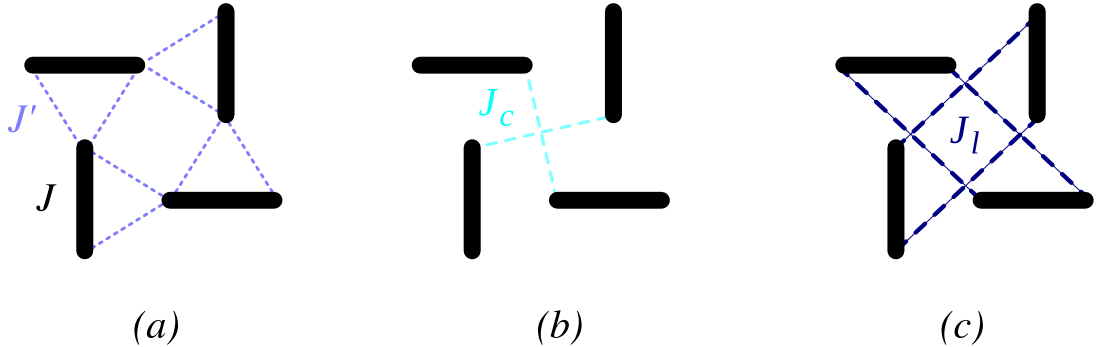}
\end{center}
\begin{minipage}[b]{0.95\textwidth}\caption{\label{fig:j7_j9}Figure illustrates the different couplings studied in this work: (a) usual couplings $J$ and $J'$ of the Shastry-Sutherland model, (b) coupling $J_{\rm c}$ creating dimerized chains along the diagonals, and (c) $J_{\rm l}$ creating two-leg ladders along the diagonals.}
\end{minipage}
\end{figure}

The total Hamiltonian studied can be written as 
\begin{equation}
 \label{eq:hamilton}
 H = H_{\rm SS} + H_{\rm c} + H_{\rm l} - B \sum_{i} S_i^z \quad ,
\end{equation}
where $H_{\rm SS}$ denotes the usual Shastry-Sutherland model with Heisenberg-type couplings $J$ and $J'$ (see Fig.~\ref{fig:j7_j9}a). The first extra term $H_{\rm c}$ represents a Heisenberg-type coupling $\propto J_{\rm c}$ which gives rise to dimerized chains in $J_{\rm c}$ along both diagonals $x\pm y$ of the Shastry-Sutherland lattice (see Fig.~\ref{fig:j7_j9}b). Similarly, the second extra term $H_{\rm l}$ is also a Heisenberg-type coupling $\propto J_{\rm l}$ which connects dimers along both diagonals $x\pm y$, but this time two-leg ladders with a rung coupling $J$ and a leg coupling $J_{\rm l}$ are formed (see Fig.~\ref{fig:j7_j9}c). Lastly, the term proportional to $B$ represents an external magnetic field in $z$-direction.   

We are actually not interested in the full phase diagram of the model. A clear hierarchy is expected for a theoretical description of SrCu$_2$(BO$_3$)$_2$: $J > J' \gg J_{\rm c}, J_{\rm l}$.  Recent ab-initio calculations estimated that $J_{\rm c}\approx\, $0.025 and $J_{\rm l} \approx\, $0.01 J\cite{mazurenko08}. One may wonder why such small couplings can have a sizable effect on the magnetization process. To this end it is important to realize that at low magnetization the relevant two-body interactions and the pure hopping start only in sixth order in $J'/J$\cite{dorier08} which is also small quantity. One finds $(J'/J)^6=0.015625$ for $J'/J=1/2$ which is indeed of the same order of magnitude. It is therefore mandatory to perform a high-order expansion in $J'/J$, but it is enough to consider only effects which are linear in $J_{\rm c}$ and $J_{\rm l}$ because higher orders are negligible. The first part has been done in Ref.~\cite{dorier08}. In this work we determine the corrections of the form $J_{\rm c} P_{\rm c} (J'/J)$ and $J_{\rm l} P_{\rm l} (J'/J)$ where $P_{\rm c}$ and $P_{\rm l}$ are polynoms in $J'/J$ which we have determined up to order 5.

\section{Method}

We have tackled Hamiltonian (\ref{eq:hamilton}) by PCUTs as we did recently for the pure Shastry-Sutherland model in a magnetic field\cite{dorier08}. We refer for a detailed introduction of the method to the existing literature\cite{knetter00,knetter03,mila08}. Here we only give the general idea and we stress certain technical aspects which arise due to the new couplings.

The main idea of the PCUT approach is to transform the initial Hamiltonian (\ref{eq:hamilton}) which changes the number of triplets to an effective Hamiltonian which conserves the number of the true quasi-particles which are triplons in our case. In PCUTs this is done efficiently to high order in perturbation. It is important to realize that the magnetic field term is unchanged during the unitary transformation because the total $S^{\rm z}_{\rm tot}$ is a conserved quantity. The relevant processes for the physics in a finite magnetic field have maximum total spin and total $S_z$. Other spin channels are relevant for spectroscopic observables which have been studied earlier for the pure Shastry-Sutherland model\cite{knetter00_2,knetter04}. The effective Hamiltonian for the physics in a finite magnetic field therefore contains only one triplon flavor with $S^{\rm z}=1$. This remaining degree of freedom is a hardcore boson\cite{momoi00,dorier08,mila08}.
 
There are several reasons why the treatment of Hamiltonian (\ref{eq:hamilton}) is more complicated than the pure Shastry-Sutherland model. First of all, there are three expansion parameters ($J'/J$, $J_{\rm c}/J$, and $J_{\rm l}/J$) compared to one ($J'/J$). Second and more important, the product state of singlets is {\it not} the exact groundstate anymore once $J_{\rm c}$ and/or $J_{\rm l}$ are finite. This results in much more intermediate states during the numerical calculation and leads consequently to a reduced maximum order which can be reached by PCUTs. Finally, the number of matrix elements of Hamiltonian (\ref{eq:hamilton}) are larger than in the pure Shastry-Sutherland model, e.g. there are only processes changing the number of triplets by one in the Shastry-Sutherland model while $H_{\rm c}$ and $H_{\rm l}$ contain both matrix elements changing the number of triplets by two. This results also in an increased effort for the PCUT calculation. 
  
\section{Effective couplings}

In this section we will present the results we have obtained for the most important processes at low magnetizations. These are on the one hand the pure kinetic hopping which is important for possible superfluid and supersolid phases and on the other hand the two-body interactions which determine the structure of the magnetization plateaux.     

\subsection{Kinetics}

\begin{figure}[h]
\begin{center}
\includegraphics[width=0.65\textwidth]{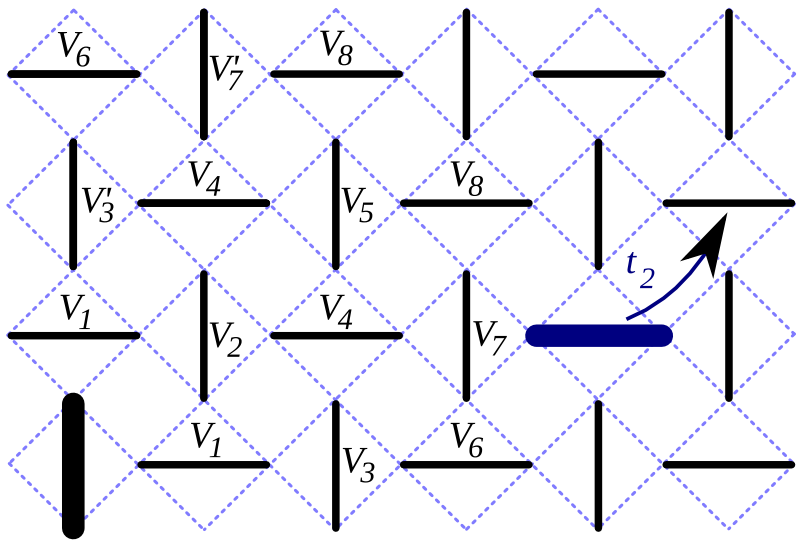}
\end{center}
\begin{minipage}[b]{0.95\textwidth}\caption{\label{fig:effective_interactions}Shastry-Sutherland lattice and definition of the 2-body interactions. $V_n$ is the coefficient of the 2-body interactions between the thick dimer and the dimer labeled $V_n$. Additionally, the definition of the hopping $t_2$ is illustrated in blue on the right.}
\end{minipage}
\end{figure}

For the pure Shastry-Sutherland model, a direct hopping of the triplon is heavily suppressed due to the strong frustration. The kinetic energy is dominated by correlated hopping processes\cite{momoi00} which favor so-called supersolid phases\cite{schmidt08}. This results in an almost flat one-triplon dispersion. The only non-vanishing hopping in the effective model up to order 15 is the hopping $t^{\rm SS}_2$ over the diagonals in the Shastry-Sutherland lattice (see Fig.~\ref{fig:effective_interactions}). As mentioned above, it only starts in sixth order with a very small coefficient $(1/96)(J'/J)^6$. Taking the full expression, one gets $t^{\rm SS}_2 / J= 0.00045$ for $J'/J=0.5$ which is a very small energy scale.

It is then obvious that the couplings $J_{\rm c}$ and $J_{\rm l}$ will dominate the pure kinetics of the triplons even if they are only a few percent of $J$. The reason is that they are contributing to $t_2$ linearly in $J_{\rm c}$ and $J_{\rm l}$. As mentioned above, the term $H_{\rm c}$ represents dimerized chains along both diagonals of the Shastry-Sutherland lattice. The corresponding matrix element for a hopping of the triplon over the diagonal is in first order $t^{\rm c}_2=-J_{\rm c}/4$. In contrast, the term $H_{\rm l}$ forms two-leg ladders over the diagonal of the Shastry-Sutherland lattice and one finds in first order $t^{\rm l}_2=J_{\rm l}/2$. Note that the different signs of the two contributions arise due to the frustration. Indeed, the coupling $J_{\rm c}$ is a frustrating diagonal coupling in the two-leg ladders formed by the couplings $J_{\rm l}$.  

In total, we obtain the following hopping over the diagonal
\begin{equation}
 \frac{t_2 }{J}  =\frac{1}{J}\left( t^{\rm SS}_2 -\frac{J_{\rm c}}{4} + \frac{J_{\rm l}}{2} \right) \quad .
\end{equation}
Assuming $J_{\rm c} \approx J_{\rm l} \approx 0.01 J$, we obtain $t_2 /J  = 0.00295$ for $J'/J=1/2$. The pure hopping has therefore increased by a factor 10 due to the extra couplings! Note that there will be no sizable effects on other hopping integrals.

\subsection{Two-body interactions}

In Ref.~\cite{dorier08}, we have studied the effective model obtained by PCUTs by analyzing the classical limit, i.e. we replaced the hardcore bosons by effective spin 1/2 using the Matsubara-Matsuda representation\cite{matsubara56} of hard-core bosons, and we treated the spins as classical vectors of length 1/2. We found that the magnetization curve is dominated by solid phases. Each solid corresponds to a phase with broken translational symmetry where triplons are frozen in a regular fashion in the ground state. These gapped phases produce plateaux in the magnetization curve. 

\begin{table}[h]

\begin{center}
\lineup
\begin{tabular}{*{8}{c}}
\br                              
$V_n$ & $J'$ & $J_{\rm c}$ & $J_{\rm l}$ &\m$V_n^{\rm SS} \left[\frac{J'}{J} = \frac{1}{2}\right]$&$V_n^{\rm J_{\rm c}} \left[\frac{J'}{J} = \frac{1}{2} , \frac{J_{\rm c}}{J} = \frac{1}{100} \right] $&$\0V_n^{\rm J_{\rm l}} \left[\frac{J'}{J} = \frac{1}{2} , \frac{J_{\rm l}}{J} = \frac{1}{100} \right] $&\cr
\mr
$V_1$ & $J'$ & $J_{\rm c} J'$ & $J_{\rm l} J'^2$ & $0.3216$ &\0 -0.0002 &\0 0.0019\vspace*{+1mm}\cr
$V_2$ & $J'^3$ & $J_{\rm c} $ & $J_{\rm l} $ & $0.0538$ &\0  0.0014 &\0 0.0048\vspace*{+1mm}\cr
$V_3$ & $J'^2$ & $J_{\rm c} J'^3 $ & $J_{\rm l} J'^4$ & $0.1862$ &\0 -0.0004 &\0 0.0001\vspace*{+1mm}\cr
$V_4$ & $J'^4$ & $ J_{\rm c} J'$ & $J_{\rm l} J'^2$ & $0.0151$ &\0 0.0011 &\0 0.0005\vspace*{+1mm}\cr
$V_{3'}$ & $J'^6$ & $ J_{\rm c} J'^3$ & $J_{\rm l} J'^4$ & $0.0034$ &\0 0.0006 &\0 0.0001\vspace*{+1mm}\cr
$V_{5}$ & $J'^6$ & $ J_{\rm c} J'^3$ & $J_{\rm l} J'^4$ & $0.0017$ &\0 0.0003 &\0 0.0000\vspace*{+1mm}\cr
$V_{7}$ & $J'^6$ & $ J_{\rm c} J'^2$ & $J_{\rm l} J'^2$ & $0.0017$ &\0 0.0009 &\0 -0.0009\vspace*{+1mm}\cr
$V_{6}$ & $J'^8$ & $ J_{\rm c} J'^4 $ & $J_{\rm l} J'^4$ & $0.0002$ &\0 0.0002 &\0 -0.0002\vspace*{+1mm}\cr
$V_{8}$ & $J'^8$ & $ J_{\rm c} J'^4$ & $J_{\rm l} J'^4$ & $0.0001$ &\0 0.0001 &\0 -0.0001\cr
\br
\end{tabular}
\end{center}
\caption{\label{tabone}This table illustrates the effect of $J_{\rm c}$ and $J_{\rm l}$ on all the considered two-triplon interactions. For a specific interaction $V_n$ (column 1), the leading order concerning the three expansion parameters $J'$, $J_{\rm c}$, and $J_l$ are given in columns 2-4. Column 5 shows the amplitude of the two-triplon interaction in the Shastry-Sutherland model for $J'/J=1/2$ using the extrapolated $15^{th}$ order series. Column 6 (column 7) gives the amplitude of the bare series linear in $J_{\rm c}$ ($J_{\rm l}$) for $J'/J=1/2$ and $J_{\rm c}/J=1/100$ ($J_{\rm l}/J=1/100$).} 

\end{table}

Surprisingly, the plateaux stabilized at low magnetizations are $1/9$, $2/15$, and $1/6$\cite{dorier08} which is in contradiction to the expected $1/8$ plateaux in the experimental compound SrCu$_2$(BO$_3$)$_2$. The classical energy of the different plateaux mainly depends on the two-triplon density-density interactions as long as we consider low magnetization. The different two-triplon interactions are illustrated in Fig.~\ref{fig:effective_interactions}. One central result of Ref.~\cite{dorier08} is that the classical energy of all stabilized structures at low magnetization (except the 1/9 plateau) depends dominantly on the three interactions $V_3'$, $V_5$, and $V_7$ which all appear in sixth order $(J'/J)^6$. This statement is also true for the 1/8 structures discussed in the literature\cite{miyahara03}. We will therefore discuss in the following the influence of $J_{\rm c}$ and $J_{\rm l}$ on all the considered two-triplon interactions but with a special focus on the interactions $V_3'$, $V_5$, and $V_7$.  
   
In Tab.~\ref{tabone} we give an overview of the effects of $J_{\rm c}$ and $J_{\rm l}$ on the two-triplon interactions. As expected, the largest interactions are not much affected by the additional couplings. These terms appear in low order in $J'/J$ and the change is only a small perturbation. But this is different for the interactions starting in higher order in $J'/J$ which are the relevant ones at low magnetization. On the right side of Tab.~\ref{tabone} we give a quantitative comparison between the interaction strength in the pure Shastry-Sutherland model at $J'/J=1/2$ (using the extrapolated $15^{\rm th}$ order series) and the additional contributions linear in $J_{\rm c}$ and $J_{\rm l}$ setting each coupling to $0.01 J$. Note that this is indeed the order of magnitude deduced by ab-initio calculations for these couplings (in fact one obtains slightly larger couplings of the order $0.025 J$ for $J_{\rm c}$)\cite{mazurenko08}. 

Among the relevant interactions for the physics at low magnetization, the coupling affected most by $J_{\rm c}$ and $J_{\rm l}$ is the two-triplon interaction $V_7$ which gets already a contribution of the order $J_{\rm c}(J'/J)^2$ and  $J_{\rm l}(J'/J)^2$. This is related to the fact that the interaction $V_3$ starts already in order $(J'/J)^2$. Indeed, the distance between the two dimers for the $V_7$ interaction is the distance for the $V_3$ interaction plus one diagonal link. The diagonal link can be done linear in $J_{\rm c,l}$ such that one finally has $J_{\rm c,l}(J'/J)^2$.

Globally, one also recognizes that the effects of $J_{\rm c}$ and $J_{\rm l}$ on the two-triplon interactions are often competing, i.e. although each change can be sizable, the combined effect of both couplings almost compensate assuming $J_{\rm c} \approx J_{\rm l}$. 

\section{Magnetization plateaux}

In this section we describe how the additional couplings $J_{\rm c}$ and $J_{\rm l}$ affect the classical energy of the relevant plateaux and its structures at low magnetization. Here we concentrate on the three structures at magnetizations $m\in\{1/9,2/15,1/6\}$ we found to be stabilized for the pure Shastry-Sutherland model\cite{dorier08} plus two structures for 1/8 plateaux discussed in the literature for SrCu$_2$(BO$_3$)$_2$\cite{miyahara00}\footnote{Note that we do not discuss in this work the possibility that the new couplings $J_{\rm c}$ and $J_{\rm l}$ induce novel solid structures or stabilize suprafluid or supersolid phases. We concentrate here on the plateaux stabilized for the pure Shastry-Sutherland model plus additional structures at $1/8$ magnetization.} The explicit expressions for the classical energy of these plateaux in terms of the two-triplon interactions $V_n$ are given in Tab.~\ref{tabtwo}. Additionally, the corresponding structures of the plateaux are illustrated.
 
\begin{table}
\begin{center}
\begin{minipage}{3.5in}
\begin{center}
\begin{tabular}{*{2}{c}}  
\br                       
Plateaux & Classical energy $[J]$\cr
\mr
$m=1/9$ & $E_{1/9}=-\frac{\mu}{9}+\frac{2 V_6 }{9}$\vspace*{+1mm}\cr
$m=1/8$ & $E^{\rm square}_{1/8}=-\frac{\mu}{8}+\frac{V_5}{4}$\vspace*{+1mm}\cr
$m=1/8$ & $E^{\rm rhomboid}_{1/8}=-\frac{\mu}{8}+\frac{V_5+V_7}{8}$\vspace*{+1mm}\cr
$m=2/15$ & $E_{2/15}=-\frac{2\mu}{15}+\frac{2V_3' +2V_7 +4 V_6  + 4V_8 }{15}$\vspace*{+1mm}\cr
$m=1/6$ & $E_{1/6}=-\frac{\mu}{6}+\frac{V_3'+2V_7}{6}$\cr
\br
\end{tabular}
\end{center}
\end{minipage}
\begin{minipage}{2.2in}
\includegraphics[width=2.2in]{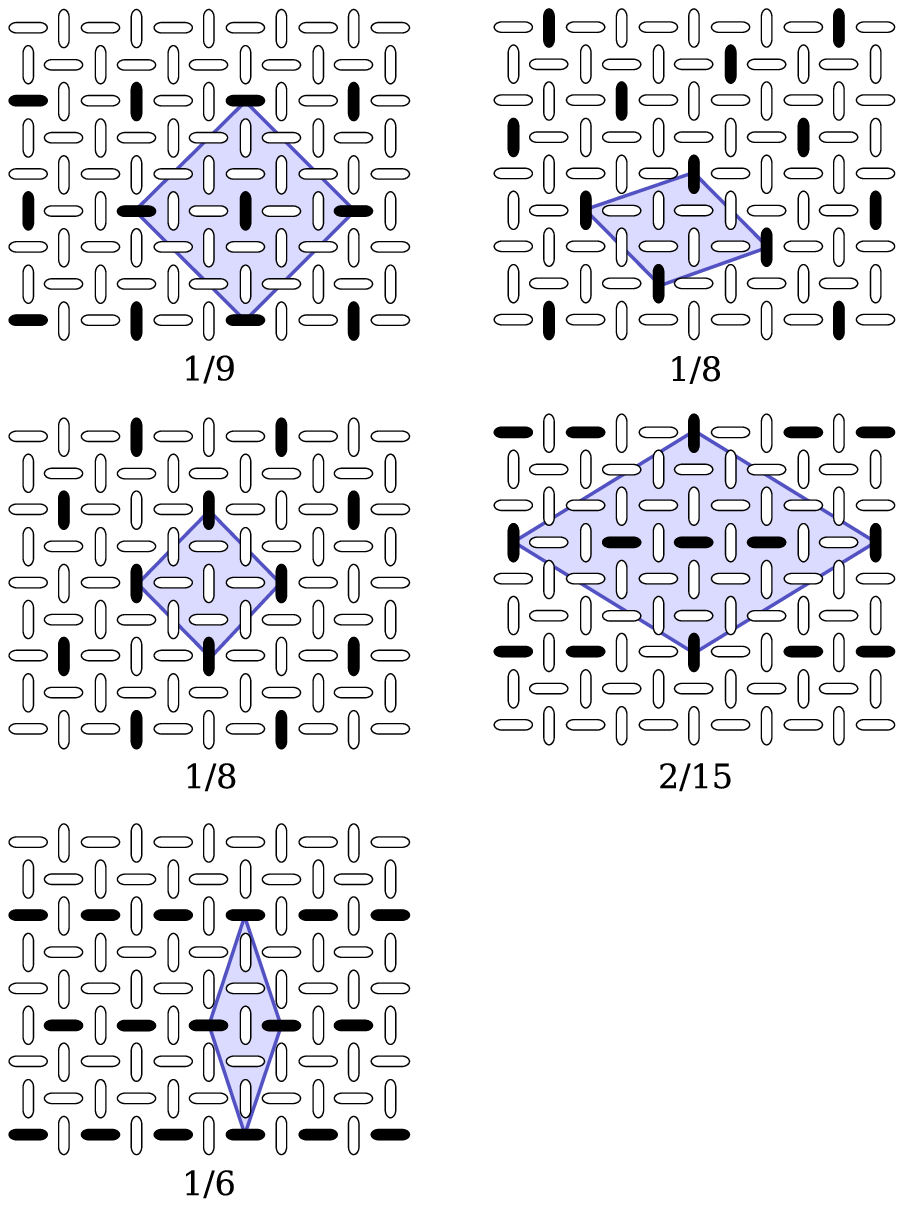}
\end{minipage}
\end{center}
\caption{\label{tabtwo}The left table shows the classical energy of the relevant plateaux at low magnetization in terms of the twp-triplon interactions. The right figure illustrates the different plateau structures and their unit cells. Black dimers denote triplets and empty dimers represent singlets.}
\end{table}

\begin{figure}[h]
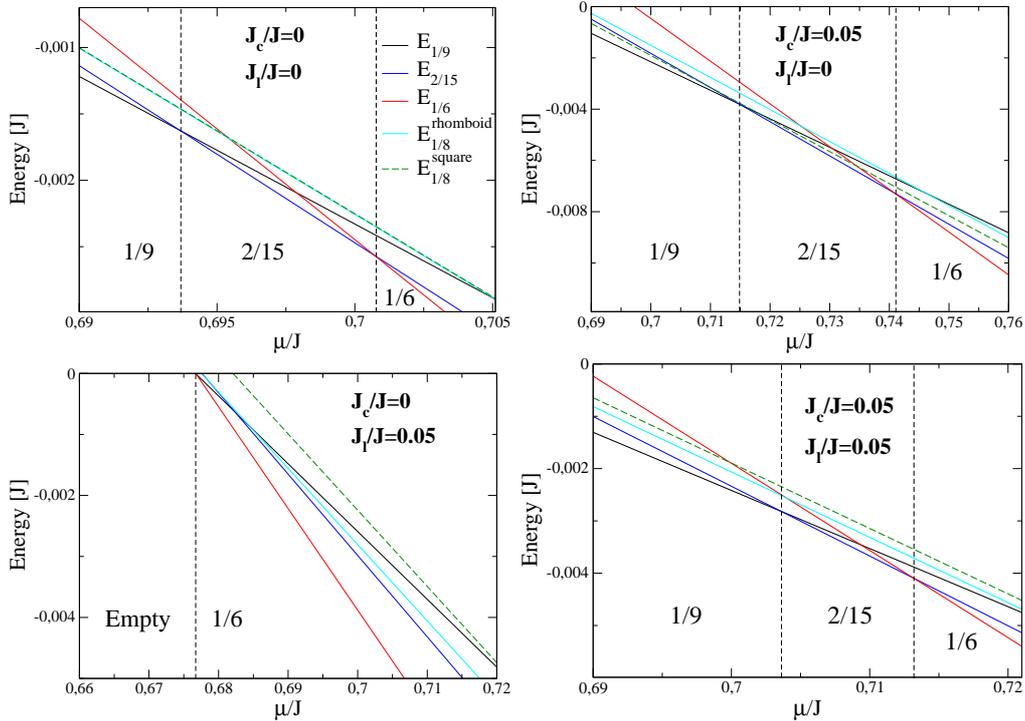

\begin{center}
\begin{minipage}[b]{2.6in}
\includegraphics[width=2.6in]{./PlateauxEnergy.Jp.0.5.J7.0.0.J9.0.0.eps}
\end{minipage}
\begin{minipage}[b]{2.6in}
\includegraphics[width=2.6in]{./PlateauxEnergy.Jp.0.5.J7.0.05.J9.0.0.eps}
\end{minipage}
\begin{minipage}[b]{2.6in}
\includegraphics[width=2.6in]{./PlateauxEnergy.Jp.0.5.J7.0.0.J9.0.05.eps}
\end{minipage}
\begin{minipage}[b]{2.6in}
\includegraphics[width=2.6in]{./PlateauxEnergy.Jp.0.5.J7.0.05.J9.0.05.eps}
\end{minipage}
\end{center}
\begin{minipage}[b]{\textwidth}\caption{\label{fig:plateaux} Classical energy of the different low-magnetization plateaux for $J'/J=1/2$ and (a) $J_{\rm c}=J_{\rm l}=0$, (b) $J_{\rm c}=0.05$ and $J_{\rm l}=0$, (c) $J_{\rm c}=0$ and $J_{\rm l}=0.05$, and (d) $J_{\rm c}=J_{\rm l}=0.05$.}
\end{minipage}
\end{figure}

In a next step we use the values for the two-triplon interactions as discussed in the last section in order to compare the energy of the different plateaux. This is done in Fig.~\ref{fig:plateaux} for different representative values of $J'$, $J_{\rm c}$ and $J_{\rm l}$. For the pure Shastry-Sutherland model (see Fig.~\ref{fig:plateaux}a), the sequence $1/9$, $2/15$, and $1/6$ is stable. 

Adding only the coupling $J_{\rm c}$, results in an increase of all the couplings $V_3'$, $V_5$, and $V_7$. Correspondingly, all classical energies are larger and the phase transitions occur at larger values of the chemical potential (see Fig.~\ref{fig:plateaux}b). The coupling affected most is the two-triplon interaction $V_7$ which affects mostly the $1/6$ plateau, then the $2/15$ plateau and also the $1/8$ rhomboid structure. One therefore obtains a larger $1/9$ plateau and also the energy of the $1/8$ square structure has only a slightly larger classical energy close to the transition from $1/9$ to $2/15$. The $J_{\rm c}$ coupling therefore tends to stabilize a $1/8$ plateau. For $J_{\rm c}>0.05$, we indeed find a stable $1/8$ plateaux with a square unit cell. Note that the $2/15$ plateau has an increased extension because the energy of $1/6$ structure has changed even more.   

We next consider the case where only the coupling $J_{\rm l}$ is turned on. This is illustrated in Fig.~\ref{fig:plateaux}c. Here the interactions $V_3'$ and $V_5$ are almost unchanged. The interaction $V_7$ is again strongly altered but this time it gets a negative contribution. So the effects are opposite to the coupling $J_{\rm c}$. The $1/6$ structure benefits most and its extension strongly increases. We would like to remark that the case of a finite $J_{\rm l}$ with vanishing $J_{\rm c}$ seems to be unrealistic since it is an exchange over a longer distance than $J_{\rm c}$.  

Finally, we consider the case where both couplings are present having the same weight. A corresponding plot is shown in Fig.~\ref{fig:plateaux}d. It should be clear from the discussion above that the effects on the interaction $V_7$ are almost compensating. One mainly has an increase of $V_3'$ and $V_5$ due to the coupling $J_{\rm c}$. Since all plateaux structures except $1/9$ depend on these two interactions, the net effect is rather small. Only the 1/9 plateau has an increased extension.

\section{Conclusions}

We have studied the magnetization process of an extended Shastry-Sutherland model. Two SU(2) invariant in-plane couplings have been added to the normal Shastry-Sutherland model. Perturbative continuous unitary transformations are used to derive the leading order effects of these couplings on the relevant hopping and interaction amplitudes. 

The most important finding of this work is that already very small couplings can affect the physics at low magnetization significantly. The physical reason is that the relevant processes in the effective model for the pure Shastry-Sutherland model in this regime start only in $6^{\rm th}$ order perturbation theory which is also a small energy scale. We found that the additional coupling $J_{\rm c}$ which is expected to be the largest in-plane correction to the Shastry-Sutherland model indeed stabilizes a $1/8$ plateau with a squared structure but the coupling strengths needed for a sizable $1/8$ plateaux are a factor 3 to 4 larger than the value deduced from first principle studies\cite{mazurenko08}. In contrast, the sum of both couplings assuming $J_{\rm c} \approx J_{\rm l}$ has only a small net effect which is a consequence of their compensating character. 

Our study clearly shows that it will be also interesting to investigate the effects of other additional couplings to the Shastry-Sutherland model like intra-plane couplings or DM interactions on the magnetization process which are expected to be of similar magnitude than $J_{\rm c}$ and $J_{\rm l}$\cite{mazurenko08,cepas01,cheng07} in order to gain a better understanding of the rich magnetization curve observed for the frustrated quantum magnet SrCu$_2$(BO$_3$)$_2$. Additionally, it will be also important to study the effect of the enhanced kinetic energy induced by the additional couplings in order to investigate whether superfluid or supersolid phases are formed. 
 
\ack

We acknowledge very useful discussions with A.~Abendschein, S.~Capponi, and G.~S.~Uhrig. KPS acknowledges ESF and EuroHorcs for funding through his EURYI. Numerical simulations were done on Greedy at EPFL ({\it greedy.epfl.ch}). This work has been supported by the SNF and by MaNEP.

\end{document}